\documentclass{svmult}
\usepackage{graphicx}
\usepackage{natbib}

\def\cm{\textrm{cm}}
\def\kpc{\textrm{kpc}}

\def\Kelv{\textrm{K}}
\def\gcm2{\textrm{g}~\textrm{cm}^{-2}}
\def\kms{\textrm{km}~\textrm{s}^{-1}}
\def\eV{\textrm{eV}}

\def\MHz{\textrm{MHz}}
\def\GHz{\textrm{GHz}}

\def\yr{\textrm{yr}}
\def\Myr{\textrm{Myr}}
\def\muGauss{\mu\textrm{G}}
\def\Msun{\textrm{M}_{\odot}}

\def\apj{ApJ}
\def\mnras{MNRAS}
\def\apjl{ApJL}
\def\nat{Nature}
\def\araa{ARA\&A}
\def\aap{A\&A}
\def\aaps{A\&AS}

\def\aj{AJ}
\def\ssr{SSRv}

\begin{document}

\title*{From 10 Kelvin to 10 TeraKelvin: Insights on the Interaction Between Cosmic Rays and Gas in Starbursts}
\author{Brian C. Lacki}
\titlerunning{From 10K to 10TK: Insights on CRs and Gas in Starbursts}
\institute{Brian Lacki \at Jansky Fellow of the National Radio Astronomy Observatory, Institute for Advanced Study, Astronomy, Einstein Drive, Princeton, NJ 08540, USA, \email{brianlacki@ias.edu}}
\maketitle

\abstract{Recent work has both illuminated and mystified our attempts to understand cosmic rays (CRs) in starburst galaxies.  I discuss my new research exploring how CRs interact with the ISM in starbursts.  Molecular clouds provide targets for CR protons to produce pionic gamma rays and ionization, but those same losses may shield the cloud interiors.  In the densest molecular clouds, gamma rays and $^{26}$Al decay can provide ionization, at rates up to those in Milky Way molecular clouds.  I then consider the free-free absorption of low frequency radio emission from starbursts, which I argue arises from many small, discrete H II regions rather than from a ``uniform slab'' of ionized gas, whereas synchrotron emission arises outside them.  Finally, noting that the hot superwind gas phase fills most of the volume of starbursts, I suggest that it has turbulent-driven magnetic fields powered by supernovae, and that this phase is where most synchrotron emission arises.  I show how such a scenario could explain the far-infrared radio correlation, in context of my previous work.  A big issue is that radio and gamma-ray observations imply CRs also must interact with dense gas.  Understanding how this happens requires a more advanced understanding of turbulence and CR propagation.}

\section{Introduction}
\label{sec:Introduction}

Starbursts are detected in synchrotron radio, which is generated from cosmic ray (CR) electrons and positrons ($e^{\pm}$) interacting with magnetic fields (\citealt{Condon92}), and now in gamma rays, which are generated by CR protons colliding with ambient gas and creating pions (\citealt{Acciari09}; \citealt{Acero09}; \citealt{Abdo10}).  While nearby starbursts can be resolved in radio emission, this is not true for the gamma rays.  Furthermore, because the gamma rays are a recent discovery, our understanding of CRs is at a basic, phenomenological level.  Thus, much of the theoretical work about CRs in starbursts uses one-zone steady-state models (\citealt{Paglione96}; \citealt{Torres04}; \citealt{Domingo05}; \citealt{deCeaDelPozo09}; \citealt{Lacki10-FRC1}).  These assume that the interstellar medium (ISM), magnetic fields, and CR injection of starbursts are completely homogeneous -- essentially, that starbursts have no structure, which is obviously not true in detail.  These kinds of models can help us understand the energetics of CRs in starbursts (which, even a few years ago, was mysterious), but the propagation is treated as a black box.

Some works have expanded from these assumptions, by correlating radio and gamma ray emission and star-formation to constrain CR diffusion (\citealt{Murphy08,Murphy12}), by creating 3D steady-state models of CR populations in the Milky Way (\citealt{Strong98}) and the nearby starbursts M82 and NGC 253 (\citealt{Persic08}; \citealt{Rephaeli10}), or by considering the time- and position- dependent contributions of individual CR accelerators to starbursts' integrated CR spectra (\citealt{Torres12}).

\begin{figure}
\centerline{\includegraphics[width=13cm]{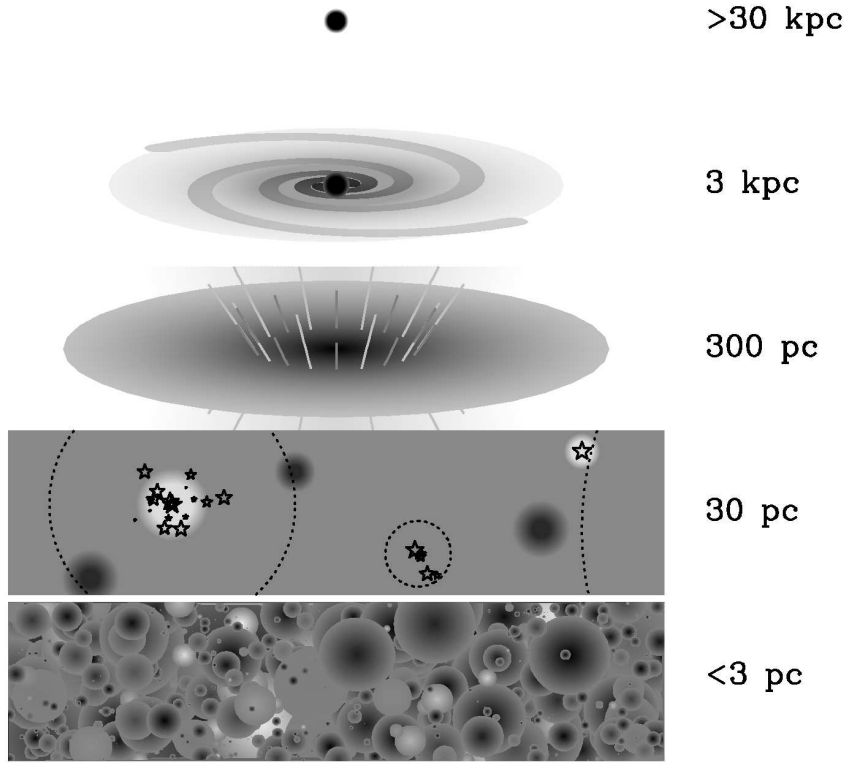}}
\caption{Starbursts appear more inhomogeneous on smaller scales (the given sizes are to order of magnitude).  At large scales ($>$ 30 kpc), the galaxy is completely unresolved (as for gamma-ray observations and high-$z$ radio observations).  Closer in, we see the resolution of the host galaxy from the starburst (3 kpc) and then the resolved starburst disk with its superwind (300 pc).  Then, individual star-forming regions and large molecular cloud complexes become apparent (30 pc).  On these scales, how CRs travel from individual accelerators becomes important (dotted circles).  Finally, on the smallest scales ($<$3 pc) it becomes clear that the ISM seethes with turbulence and is extremely complex.  The overall energetics of CRs in starbursts can be represented by the large-scale view of simple one-zone models.  More advanced models take a finer view, simulating large-scale density gradients or the effects of individual CR accelerator regions.  Yet, the physics of the CR propagation depends on the chaos of the multiphase ISM on small scales, and so is very poorly understood.\label{fig:Scales}}
\end{figure}

Yet even these approaches are simplifications of what actually happens in starbursts.  On small enough scales, the apparently smooth fluctuations in gas density and star-formation resolve into a highly chaotic froth with many phases of vastly different physical conditions (Figure~\ref{fig:Scales}).  Molecular gas is highly turbulent in starbursts, leading to strong density fluctuations (\citealt{Krumholz05}).  In star-forming regions, the overdensities of the molecular gas collapse still further into molecular cores and ultimately into protostars.  Newborn O stars and star clusters carve out H II regions.  Supernovae and clusters blast bubbles of $10^7 - 10^8$ K plasma which can ultimately fill the volume of a starburst and form into a galactic wind (\citealt{Chevalier85}; \citealt{Strickland09}).  Pervading these phases are magnetic fields and the CRs themselves.  

The interaction between CRs and these different phases must be understood if we are to interpret the results of our phenomenological models.  Which phase has the magnetic fields responsible for the radio emission?  Which phase provides the target atoms responsible for pionic gamma ray emission?  Which phase regulates how CRs are transported spatially?  And how do the CRs shape these phases, if at all?

\section{How Molecular Clouds Can Stop Cosmic Rays and the Role of Gamma Rays and $^{26}$Al}
CRs are thought to interact with the bulk molecular ISM, containing most of the gas mass of starbursts, and with some good reason.  The observations of gamma rays from starburst galaxies provide evidence that CRs spend some time in dense (presumably, molecular) gas.  Another line of evidence comes from the observed Zeeman splitting of OH lines in Ultraluminous Infrared Galaxies (ULIRGs), which imply that milliGauss magnetic fields are present in OH megamaser regions (\citealt{Robishaw08}).  These regions occur in dense molecular gas, where the number density is at least $10^4\ \cm^{-3}$ and likely $10^{6 - 7}\ \cm^{-3}$ (\citealt{Robishaw08}).  However, magnetic fields require free charges to be frozen into, otherwise the magnetic field lines would slip off the gas through ambipolar diffusion (e.g., \citealt{Mestel56}).  Therefore, we have evidence that something is ionizing the dense molecular gas in ULIRGs.  Since UV light is quickly absorbed and AGNs are not always nearby to provide X-rays, the default assumption is that CRs are ionizing this gas.

The high energy densities of CRs in starbursts can sustain high ionization rates ($\zeta \approx 10^{-16} - 10^{-14}\ \sec^{-1}$) and heat the gas to high temperatures ($\sim 50 - 100\ \Kelv$) (\citealt{Suchkov93}; \citealt{Papadopoulos10-CRDRs}).  The molecular gas in starbursts may represent a ``Cosmic Ray Dominated Region'' (CRDR).  It is not clear, though, how much of the gas the CRs can reach.  The problems that face ionizing radiation can be understood by considering other kinds of radiation.  Extreme ultraviolet photons interact very strongly with the gas, surrendering much of their power into ionization.  Yet that same interaction prevents the photons from going very far: they instead create small highly-ionized H II regions, leaving the majority of the starburst untouched.  On the other hand, we never consider Neutrino Dominated Regions, even though neutrinos can penetrate any reasonable column and fill the entire starburst volume.  They interact so weakly that their power just escapes the starburst rather than contributing to ionization.  To ionize a parcel of gas evenly, we must pick some kind of radiation such that the gas' absorption optical depth is $\sim 1$, and this applies to CRs.  If CRs interact weakly with the gas, they are blown out by the starburst wind before they can ionize or heat it, or they simply escape through diffusion.  If CRs interact strongly with the gas, they lose energy to pionic and ionization losses (``proton calorimetry'', \citealt{Thompson07}), stopping them and shielding the densest regions.

The gamma-ray observations of M82 and NGC 253 indicate that CR protons lose perhaps $\sim 30 - 40\%$ of their energy to pionic losses before escaping, implying much stronger losses than in the Milky Way (\citealt{Lacki11-Obs,Abramowski12,Ackermann12}).  Moreover, the hard GeV-TeV gamma-ray spectra indicate advective losses which simply remove CRs from the starburst, rather than a high diffusion constant that would let CRs penetrate deep into molecular clouds (e.g., \citealt{Abramowski12}).

\begin{figure}
\centerline{\includegraphics[width=8cm]{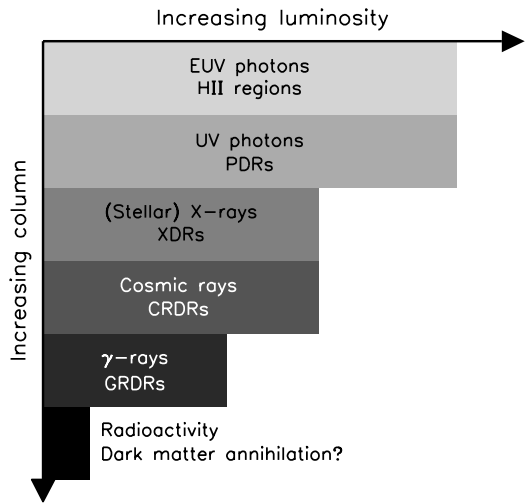}}
\caption{Different kinds of ionizing radiation are able to penetrate through different columns in the ISM, creating a hierarchy of ionization regions.  Starbursts emit many ultraviolet photons but these are stopped very rapidly, resulting small highly-ionized regions.  The injected power from gamma rays and radioactivity is small, but it can sustain a low level of ionization everywhere in the ISM. \label{fig:IonHierarchy}}
\end{figure}

The advantage -- and trouble -- of CRs is that they are deflected by magnetic fields.  On the one hand, the effective scattering of CRs lets them build up to large energy densities.  On the other, this also means that CRs see a much bigger column than a neutral particle would, stopping them much closer to their sources. The reach of CR ionization thus depends strongly on propagation.  But if CRs are not guaranteed to reach all of the molecular gas, we need something else that can.

There \emph{is} a known source of neutral radiation that can ionize the densest molecular gas in starbursts: gamma rays.  Above a few MeV, they are stopped by $\gamma Z$ pair production over columns of $\sim 200\ \gcm2$ (\citealt{Berestetskii79}).  The pair $e^{\pm}$ then ionize the gas.  Hence, even if CRs are quickly stopped, the bulk gas of the starburst is then likely to be a Gamma Ray Dominated Region (GRDR; Figure~\ref{fig:IonHierarchy}).  Gamma ray ionization harnesses proton calorimetry: the more strongly the CRs interact with the gas, the higher the energy density of gamma rays is.  The gamma-ray ionization rate is
\begin{equation}
\zeta_{\gamma} \approx F_{\gamma} \sigma_{\gamma Z} f_{\rm ion}^{\gamma} / E_{\rm ion}
\end{equation}
where $F_{\gamma}$ is the gamma-ray flux within the starburst, $\sigma_{\gamma Z} \approx 10^{-26}\ \cm^{2}$ is the cross section for $\gamma Z$ pair production (\citealt{Berestetskii79}), $f_{\rm ion}^{\gamma}$ is the efficiency that energy in gamma rays goes into ionizing gas (I assume 0.1, supported by a cascade calculation), and $E_{\rm ion} = 30\ \eV$ (\citealt{Cravens78}) is the energy used in each ionization of an atom.  $F_{\gamma}$ depends on the surface density of star-formation and how proton calorimetric the starburst is -- gamma-ray ionization is most effective in very compact ULIRGs like Arp 220.

\begin{figure}
\centerline{\includegraphics[width=8cm]{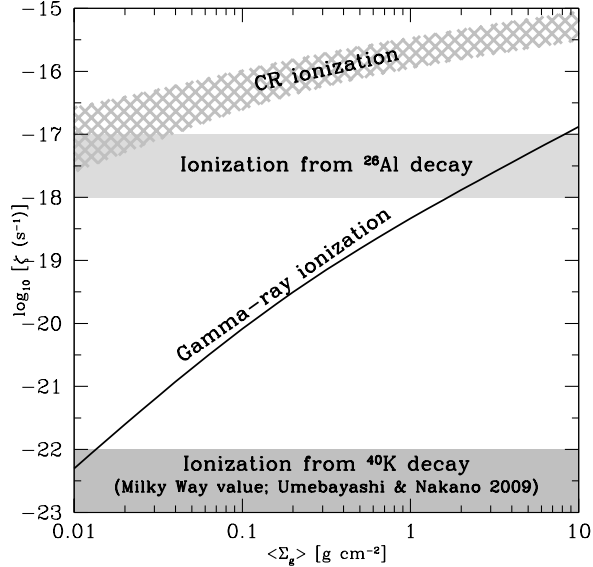}}
\caption{Comparison of ionization rates from CRs, gamma rays, $^{26}$Al, and $^{40}$K for starbursts on the Schmidt Law.  The CR and $^{26}$ Al ionization rates depend on their propagation.\label{fig:IonizationRates}}
\end{figure}

By using the Schmidt Law (\citealt{Kennicutt98}) to relate gas density to star-formation surface density, I calculate the gamma-ray ionization rate in starbursts.  These rates are plotted in Figure~\ref{fig:IonizationRates}: they range from $10^{-22}\ \sec^{-1}$ at low densities to $10^{-17}\ \sec^{-1}$.  At the highest gas surface densities, the ionization rates from gamma rays alone are comparable to those sustained by CRs in Milky Way molecular clouds (\citealt{Lacki12-GRDRs}).  Unlike CRDRs, though, GRDRs are not substantially heated by the high energy ionizing radiation.  The power dumped by gamma rays is never enough to heat the gas more than a few Kelvin from absolute zero; this is overwhelmed by dust heating, which raises the gas temperature to $\sim 50\ \Kelv$.  The coldness of the GRDRs provides a potential observational signature.  In molecular lines, the cold GRDRs will appear as shadows against the brighter, hotter CRDR material; the GRDRs should also have a narrower thermal line width (\citealt{Lacki12-GRDRs}).  

While the gamma-ray ionization rate is high in starbursts like Arp 220, is there anything that can enhance ionization where CRs cannot penetrate in less extreme starbursts?  The decay of short-lived radionuclides like $^{26}$Al inject MeV particles, ionizing the ISM (long-lived isotopes like $^{40}$K sustain only $\zeta \approx 10^{-22}\ \sec^{-1}$) (\citealt{Umebayashi09}).  The mean abundance $X$ of a radionuclide in the ISM is proportional to the star-formation rate divided by the gas mass, inversely proportional to the gas consumption time.  Quicker gas consumption means that more of the gas is converted into stars and $^{26}$Al per decay time:
\begin{equation}
X(^{26}{\rm Al}) = \frac{M(^{26}{\rm Al})}{M_{\rm gas}} = \frac{\dot{M}(^{26}{\rm Al}) \tau_{\rm decay}}{M_{\rm gas}} \propto \frac{\rm SFR}{M_{\rm gas}} = \frac{1}{\tau_{\rm gas}}.
\end{equation}
The gas consumption time in starbursts, $\sim 20\ \Myr$ in M82 (using the gas mass from \citealt{Weiss01}), is $\sim 100$ times shorter than in the Milky Way (\citealt{Diehl06}), implying mean $^{26}$Al abundances in starbursts are $\sim 100$ times higher (\citealt{Lacki12-Al26}).  The $^{26}$Al alone sustains ionization rates of $10^{-18}\ \sec^{-1} - 10^{-17}\ \sec^{-1}$.  This suggests that $^{26}$Al decay ionizes the dense gas in weak starbursts, and gamma rays ionize it in strong starbursts (Figure~\ref{fig:IonizationRates}).  

However, if it takes more than a few Myr for the $^{26}$Al to mix with the starburst molecular gas, it cannot contribute to ionization (\citealt{Meyer00}).  Furthermore, the $^{26}$Al may be dumped into the superwind and advected away instead of ionizing molecular gas.  On the other hand, the distribution of 1.809 MeV gamma rays from the decay of $^{26}$Al in our own Galaxy suggests that it is formed by very massive stars (\citealt{Knoedlseder99}).  These very massive stars will not have time to migrate from their star-formation regions, and will likely be near star-forming molecular gas.  Like CRs, then, the ionization rate from $^{26}$Al nuclei depends on the unknown details of their propagation, something which does not affect gamma rays.

Thus, CRs do not necessarily reach all of the molecular gas in the starbursts.  Depending on the unknown propagation of CRs, it is conceivable at least that they leave most of the molecular gas untouched.  Ionization from gamma rays and radioactivity may penetrate deeper columns, creating the required free charges to retain magnetic fields.  But if CRs are not thoroughly mixed with the molecular gas, are they merely confined to pockets of that gas, or do they spend most of their time in another ISM phase entirely?

\section{H II Regions and Starbursts' MHz Radio Spectra: Str\"omgren Sphere Pudding}
The molecular gas ultimately collapses into stars: among the youngest stars are massive O stars that generate ionizing photons.  The ionizing luminosity in Lyman continuum photons is actually much greater than the luminosity in CRs; however, these ultraviolet photons interact so strongly with the gas that they never travel far from their birth stars.  Thus, instead of the weak but pervasive ionization of CRs or gamma rays, the ultraviolet photons create fully ionized H II regions surrounding O stars with low filling factor (Figure~\ref{fig:IonHierarchy}).  

\begin{figure}
\centerline{\includegraphics[width=12cm]{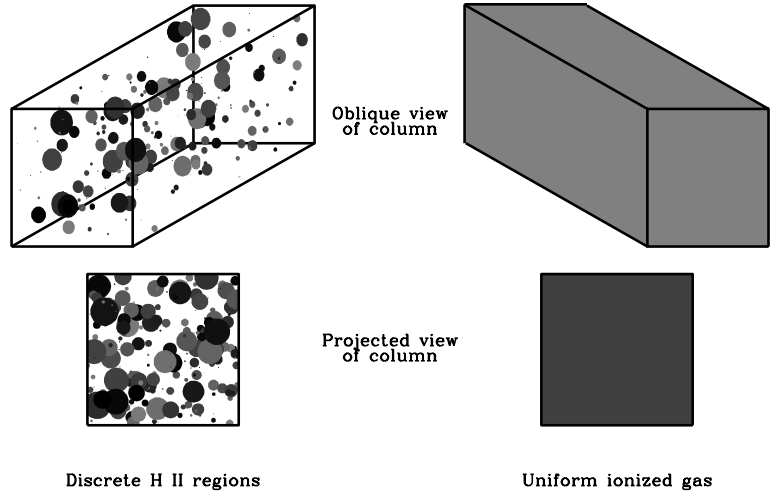}}
\caption{Illustration of the difference between a uniform density collection of H II regions and a truly uniform ionized gas.  In each case, there is free-free absorbing ionized gas which is spread throughout the starburst, so a ``uniform slab'' model is useful.  As the frequency drops to zero, the ionized gas becomes completely opaque.  However, not every sightline passes through an H II region on the left, as seen by the white space in the face-on view: the starburst is partially uncovered.  Furthermore, even on sightlines intersecting an H II regions, that ionized gas is at some depth within the starburst (as seen in the oblique view).  Thus the H II regions block only a constant fraction of radio emission at low frequency, unlike a uniform gas. \label{fig:HIIExplanation}}
\end{figure}

The H II regions have two effects on the observed radio spectra: they make the free-free emission observed at high frequencies, and they host free-free absorption at low frequencies (\citealt{Condon92}).  Free-free absorption from individual H II regions is actually observed at low frequencies, both towards the Galactic Center (e.g., \citealt{Nord06}) and in the M82 starburst (\citealt{Wills97}).  The free-free emission does not depend much on the geometry of the H II regions. This has led to the adoption of the ``uniform slab'' assumption, where a uniform density of ionized gas pervades the starburst.   However, the geometry matters for calculating the free-free absorption: in reality we have a uniform density, not of \emph{gas}, but of \emph{H II regions}.  If the density of H II regions is low enough, on some sightlines there may be no absorption: the H II regions partially cover the starburst (Figure~\ref{fig:HIIExplanation}).  Furthermore, the nearest H II region on any sightline is buried part way into it.  Thus, \emph{if} the synchrotron emitting material is outside the H II regions, some of it remains unobscured at any frequency.  In contrast, the radio flux steeply falls towards low frequencies in a truly uniform slab of ionized gas.

\begin{figure}
\centerline{\includegraphics[width=8cm]{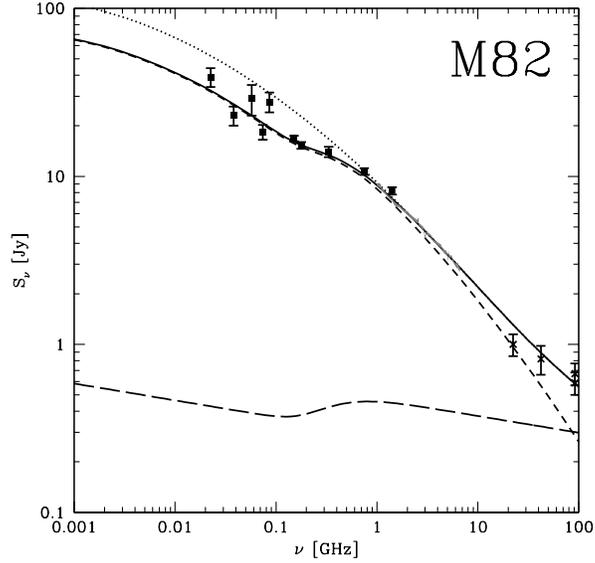}}
\caption{My fit to the 22 MHz - 92 GHz radio spectrum of M82, which assumes that free-free absorption comes from discrete H II regions around super star clusters with O stars.  The observed flux is the solid line and the unabsorbed flux is the dotted line.  I also break the observed flux down into its synchrotron (short-dashed) and thermal (long-dashed) components.  Radio data compiled in \citet{Lacki12-LowNu} (from \citealt{Roger86}, \citealt{Klein88}, \citealt{Israel90}, \citealt{Cohen07}, \citealt{Laing80}, \citealt{Hales91}, \citealt{Basu12}, and \citealt{Williams10}).  \label{fig:M82LowNu}}
\end{figure}

The effective absorption coefficient from H II regions is their number density times an effective cross section that depends on their size and opacity.  The H II regions can be thought of as Str\"omgren spheres around super star clusters.  I assume these clusters emit ionizing photons for a time $t_{\rm ion}$ in proportion to their mass $M$ and then shut off.  Using a super star cluster mass function of $dN/dM \propto M^{-2} \exp(-M / 5 \times 10^6\ \Msun)$ for $M \ge 1000\ \Msun$ (e.g., \citealt{McCrady07}), I find an effective absorption coefficient of 
\begin{equation}
\alpha_{\rm eff} = (120\ \kpc)^{-1} \left(\frac{\rho_{\rm SFR}}{\Msun\ \yr^{-1}\ \kpc^{-3}}\right) \left(\frac{n_H}{1000\ \cm^{-3}}\right)^{-4/3} \left(\frac{t_{\rm ion}}{10\ \Myr}\right)
\end{equation}
at low frequencies.  The H II regions around these SSCs become opaque at a frequency
\begin{equation}
\nu_{\rm ff}^{\rm SSC} = 700\ \MHz \left(\frac{M}{1000\ \Msun}\right)^{1/6} \left(\frac{T}{10^4\ \Kelv}\right)^{-3/4} \left(\frac{n_H}{1000\ \cm^{-3}}\right)^{2/3};
\end{equation}
the free-free absorption feature appears in the starburst spectrum around this frequency.  If I instead assume that the free-free absorption comes from H II regions around individual O stars with ionizing photon luminosities of $10^{49}\ \sec^{-1}$, then $\alpha_{\rm eff}$ is $\sim 4$ times higher.  This quantity can then be used in the uniform slab solution to model the actual amount of free-free absorption (\citealt{Lacki12-LowNu}).  I find this effective absorption coefficient is $\alpha_{\rm eff} \approx (0.1 - 1\ \kpc)^{-1}$ as $\nu \to 0$ in M82.  

In Figure~\ref{fig:M82LowNu}, I model the radio spectrum of M82 by assuming the free-free absorption and emission comes from Str\"omgren sphere H II regions around super star clusters.  I fit to the number of H II regions, their temperature, and density, as well as the unabsorbed synchrotron spectrum shape using the \citet{Williams10} parameterization:
\begin{equation}
\log_{10} \left(\frac{S_{\rm nt}^{\rm unabs}}{\rm Jy}\right) = {\cal A} + {\cal B} \log_{10} \left(\frac{\nu}{\rm GHz}\right) + {\cal C} \log_{10} \left(\frac{\nu}{\rm GHz}\right)^2
\end{equation}
and the amount of free-free emission (including both the free-free emission from the H II regions responsible for the absorption, and a fit component of additional free-free emission):
\begin{equation}
S_{\rm ff}^{\rm unabs} = {\cal D} \left(\frac{\nu}{\rm GHz}\right)^{-0.1}.
\end{equation}
The best-fit model has a spectral index ${\cal B} = -0.56$ and an intrinsic synchrotron spectral curvature of ${\cal C} = -0.08$.  Thermal emission is just 5\% of the 1 GHz flux, even including free-free emission not associated with the absorption.  The covering fraction of the H II regions is only 50\%.  In this model, the absorbed flux is still 62\% of the unabsorbed flux at low frequencies (\citealt{Lacki12-LowNu}), unlike previous models, where the synchrotron flux drops precipitously below 1 GHz.  

I should note, however, that this fit is preliminary, since the beam sizes of the radio data I used in the fit vary widely.  Some of the data, particularly the 22.5 and 38 MHz points have extremely large beam sizes ($\sim 1^{\circ}$), and could easily be contaminated, for example, by the host galaxy of M82.  However, the 74 MHz Very Large Array (VLA) sky survey resolved the inner arcminute (\citealt{Cohen07}), and the 333 MHz observations by the Giant Metrewave Radio Telescope (GMRT) actually resolved the starburst itself (\citealt{Basu12}).  Both data points are approximately in line with the other data, suggesting that the starburst itself actually is seen at low frequencies.  When I redo the fit only including data points with beam sizes smaller than 10 arcminutes, the basic conclusion that most of the radio flux is transmitted stands.  It would be interesting to measure the spectrum of the starburst with low frequency interferometers like the upgraded Jansky VLA, the GMRT, or the Low Frequency Array (LOFAR).

H II regions naively seem like a good candidate for the location of the CRs: their densities are at least the mean density of the starburst (\citealt{RodriguezRico04}), consistent with models; they surround the star-forming regions that may accelerate CRs; and as highly ionized dense plasmas, they have low Alfven velocities, so CRs can self-confine effectively (\citealt{Kulsrud69}).  However, the strong radio emission that seems to be coming from M82 at low radio frequencies constrains this hypothesis -- CRs spend some time outside the H II regions.  

\section{Supernova-Driven Magnetic Fields in the Hot Wind Phase: The Secret of the FIR-Radio Correlation?}
So where do CRs spend most of their time?  The resolved radio emission appears to be mostly diffuse.  If we take that at face value, CRs and the synchrotron emission really do fill most of the starburst.  Then the CRs reside in the phase that fills most of the starburst volume, which is thought to be the hot ($\sim 10^7 - 10^8\ \Kelv$) phase that ultimately forms into the wind (\citealt{Heckman90}).  Evidence for this gas comes from the observations of 6.7 keV iron line emission from M82; the softer diffuse X-ray emission is mostly thought to be low filling factor gas resulting from when the wind crashes into the galactic halo (\citealt{Strickland07}).  This is supported by the evidence for CR advection in M82 and NGC 253 (\citealt{Lacki11-Obs,Abramowski12}).  The picture that arises is that CRs in starbursts are similar to CRs in normal galaxies: the CRs spend most of their time in low density gas that occupies more volume but occasionally diffuse through the high density molecular gas.  Pionic gamma rays and secondary $e^{\pm}$ arise during the brief times when the CRs plunge into the molecular gas, but the synchrotron emission arises during the intervals when CRs traverse the diffuse hot phase.  Radio emission then mostly traces \emph{magnetic fields in the hot phase}.

What are these fields?  The magnetic fields of starbursts are thought to be driven by turbulent dynamos.  The energy of supersonic turbulence is injected at some outer scale $\ell$, but it rapidly dissipates over a timescale $\ell / \sigma$, where $\sigma$ is the typical velocity dispersion of the turbulence (\citealt{Stone98}):
\begin{equation}
U_{\rm turb} = \frac{1}{2} \rho \sigma^2 \approx \frac{\epsilon \ell}{\sigma},
\end{equation}
where $\rho$ is the gas density and $\epsilon$ is the volumetric energy injection rate.  The energy density is then equal to the volumetric energy injection times the dissipation time.  I suppose the outer scale of turbulence is the scale height of the starburst, the largest plausible size.  Using the central superwind density from \citealt{Strickland09}, I then find $\sigma \approx 2500\ \kms$, indicating weakly supersonic turbulence (for a sound speed $c_S \approx 1500\ \kms$).  The magnetic fields are likely amplified until they are in equipartition with turbulent energy (\citealt{Stone98}; \citealt{Groves03}).  I find for galaxies on the Schmidt law (\citealt{Kennicutt98}), using the energy injection rate from \citet{Strickland09},
\begin{equation}
B_{\rm turb}^{\rm hot} \approx 60\varepsilon\ \muGauss  \left(\frac{\Sigma_{\rm SFR}}{\Msun\ \yr^{-1}\ \kpc^{-2}}\right)^{1/2} \approx 350\varepsilon \ \muGauss \left(\frac{\Sigma_g}{\gcm2}\right)^{0.7}.
\end{equation}
for star-formation and gas surface densities $\Sigma_{\rm SFR}$ and $\Sigma_g$ respectively, and where $\varepsilon$ is a factor of order unity (\citealt{Lacki12-FRCDerived}). 

The radio emission of star-forming galaxies of star-forming galaxies is tightly correlated with its far-infrared (FIR) emission, strongly constraining the effective magnetic fields.  In \citealt{Lacki10-FRC1}, we found a phenomenological scaling of $B_{\rm FRC} \approx 400\ \muGauss (\Sigma_g / \gcm2)^{0.7}$ was necessary for this FIR-radio correlation to work.  The $B \propto \Sigma_g^{0.7}$ scaling is needed because (1) it keeps the radiation and magnetic fields in equipartition, so the ratio of Inverse Compton and synchrotron cooling times remain constant (\citealt{Volk89}):
\begin{equation}
\frac{t_{\rm synch}}{t_{\rm IC}} = \frac{U_B}{U_{\rm IC}} \approx \frac{B^2 / (8 \pi)}{F_{\rm rad} / c} \propto \frac{B^2}{\Sigma_{\rm SFR}} \propto \left(\frac{B}{\Sigma_g^{0.7}}\right)^2,
\end{equation}
using the \citet{Kennicutt98} Schmidt law ($\Sigma_{\rm SFR} \propto \Sigma_g^{1.4}$).  (2) At constant synchrotron frequency (instead of constant electron energy), it preserves the ratio of bremsstrahlung and synchrotron cooling times.  The characteristic synchrotron frequency of an electron with energy $E_e$ and charge $e$ is
\begin{equation}
\nu_C = \frac{3 E_e^2 e B}{16 m_e^3 c^5} = 1.3\ \GHz \left(\frac{E_e}{\rm GeV}\right)^2 \left(\frac{B}{100\ \muGauss}\right)
\end{equation}
(\citealt{Rybicki79}).  Plugging that result into formulae for the synchrotron (\citealt{Rybicki79}) and bremsstrahlung (\citealt{Strong98}) lifetimes give:
\begin{eqnarray}
t_{\rm synch} & \approx & 1.4\ {\rm Myr}\ \left(\frac{B}{100\ \muGauss}\right)^{-3/2} \left(\frac{\nu}{\GHz}\right)^{-1/2}\\
t_{\rm brems} & \approx & 310\ {\rm kyr}\ \left(\frac{n_H}{100\ \cm^{-3}}\right)^{-1}\\
\frac{t_{\rm synch}}{t_{\rm brems}} & \approx & 4.5 \left(\frac{B}{100\ \muGauss}\right)^{-3/2} \left(\frac{\nu}{\GHz}\right)^{-1/2} \left(\frac{n_H}{100\ \cm^{-3}}\right) \\
& \propto & B^{-3/2} \nu^{-1/2} n_H.
\end{eqnarray}
For a given gas scale height, this means that $t_{\rm synch} / t_{\rm brems} \propto (\Sigma_g^{0.7})^{-3/2} \Sigma_g \propto \Sigma_g^{-0.05}$.  The $B_{\rm turb}^{\rm hot}$ I quote satisfies these conditions.  This suggests that the FIR radio correlation is ultimately powered by the Schmidt Law, causing the hot phase turbulent magnetic fields to have the necessary scaling with mean gas density (\citealt{Lacki12-FRCDerived}).  

This model has interesting implications for whether CRs and magnetic fields are in equipartition with magnetic fields in starbursts.  Both CRs and magnetic fields would ultimately have the same power source, the mechanical energy of supernovae.  Furthermore, the fraction of supernova power they receive is likely similar (of order tens of percent), with turbulence perhaps receiving a few times more power than the CRs.  When the CRs are removed by advection (as seems to be the case in M82 and NGC 253), the residence time for both the CRs and turbulence are also similar.  In each case, the residence time is of order a sound-crossing time, with the CR lifetime perhaps being a few times longer than $\ell / \sigma$, because the wind takes some time to accelerate to $c_S$ and beyond.  Then equipartition holds between magnetic fields and CRs, even if there is no direct coupling between the two.  However, in strongly proton calorimetric galaxies, the residence time of CRs is instead set by pionic loss times that are much shorter than the sound-crossing time: the CR energy density becomes smaller than the magnetic energy density (compare with \citealt{Lacki10-FRC1}).  

Equipartition seems to work in M82 and NGC 253, as suggested by one-zone models (\citealt{Persic08,Rephaeli10}) and gamma-ray observations (\citealt{Persic12,Abramowski12}).  However, in Arp 220 and other ULIRGs, equipartition fails if the far-infrared radio correlation is to work (\citealt{Condon91,Thompson06}), and one-zone models of the radio emission indicate stronger magnetic fields, a few milliGauss (\citealt{Torres04}), than expected from equipartition.  It will be interesting to see if equipartition actually fails in Arp 220 if it is detected in gamma rays, perhaps with the Cherenkov Telescope Array (CTA; \citealt{Actis11}).

While this can explain the magnetic fields, what about the gas density?  Previous one-zone models to interpret starbursts' nonthermal emission assume that CRs experience the \emph{mean} (volume-averaged) gas density in the starburst.  This can only happen if the CRs leave the hot phase and enter the molecular gas (or possibly H II regions).  Interaction with dense gas is needed to flatten the radio spectra through bremsstrahlung and ionization losses (\citealt{Thompson06}), though advective escape also flattens spectra.  Finally, much of starbursts' radio emission probably comes from pionic $e^{\pm}$ (\citealt{Rengarajan05}; \citealt{Lacki10-FRC1}): these are created in dense gas, but they would have to diffuse out into the hot phase to emit radio.  How does this mixing of CRs with gas work?  Observations of the Galactic Center suggest that CRs \emph{don't} mix with the dense molecular gas there (\citealt{Crocker11-Wild}), but M82 and NGC 253 seem to be much more efficient at stopping protons (\citealt{Lacki11-Obs,Abramowski12}).  And why does the assumption of mean density seem to work?  Only knowledge of how CRs traverse the different phases can answer these questions.

\section{Conclusion}
The matter in a starburst ISM spans the range of energies from 10 Kelvin to 10 teraKelvin: cold molecular gas at 10 Kelvin, H II regions at 10 kiloKelvin, the hot wind phase at 10 megaKelvin, the MeV ionizing particles from $^{26}$Al decay at 10 gigaKelvin, and the cosmic rays and gamma rays at 10 teraKelvin and beyond.  The relationship between these phases, and their effects on observables, is complex.

My hypothesis here is that CRs in starbursts spend most of their time in the volume-filling hot phase of the ISM, but they occasionally mix into the molecular gas to efficiently produce gamma rays and pionic $e^{\pm}$.  The relatively efficient pionic losses within the molecular phase may shield the densest gas.  How this works in detail, like how the competition between the advection in the hot phase and the pionic losses in the molecular gas somehow lets us just use the average gas density of starbursts in one-zone models, is not clear.  

Yet the ideas presented here are still simplistic compared to the reality of gas in starbursts.  I have considered each phase as if it had a single density and magnetic field, but turbulence creates fluctuations in both of these quantities.  Ultimately, to understand CRs in starbursts, we must learn what turbulence does to CRs, gas, and magnetic fields as they interact with each other.

\subsection*{Acknowledgements}
I am supported by a Jansky Fellowship from the National Radio Astronomy Observatory.  The National Radio Astronomy Observatory is operated by Associated Universities, Inc., under cooperative agreement with the National Science Foundation.  I would like to acknowledge discussions and comments with Todd Thompson, Padelis Papadopolous, Jim Condon, Rainer Beck, and Diego Torres.


\begin{thebibliography}{}

\bibitem[Abdo et al.(2010)]{Abdo10} Abdo, A.~A., et al.\ 2010a, \apjl, 709, L152 

\bibitem[Abramowski et al.(2012)]{Abramowski12} Abramowski, A., et al.\ 2012, arXiv:1205.5485 

\bibitem[Acciari et al.(2009)]{Acciari09} Acciari, V.~A., et al.\ 2009, \nat, 462, 770 

\bibitem[Acero et al.(2009)]{Acero09} Acero, F., et al.\ 2009, Science, 326, 1080 

\bibitem[Ackermann et al.(2012)]{Ackermann12} Ackermann, M., et al.\ 2012, arXiv:1206.1346 

\bibitem[Actis et al.(2011)]{Actis11} Actis, M., Agnetta, G., Aharonian, F., et al.\ 2011, Experimental Astronomy, 32, 193 

\bibitem[Basu et al.(2012)]{Basu12} Basu, A., Mitra, D., Wadadekar, Y., \& Ishwara-Chandra, C.~H.\ 2012, \mnras, 419, 1136 

\bibitem[Berestetskii et al.(1979)]{Berestetskii79} Berestetskii, V. B., Lifshitz, E. M., \& Pitaevskii, L. P. 1979, \emph{Quantum Electrodynamics}, 2nd ed. (Oxford: Butterworth-Heinemann)

\bibitem[Chevalier \& Clegg(1985)]{Chevalier85} Chevalier, R.~A., \& Clegg, A.~W.\ 1985, \nat, 317, 44

\bibitem[Cohen et al.(2007)]{Cohen07} Cohen, A.~S., Lane, W.~M., Cotton, W.~D., et al.\ 2007, \aj, 134, 1245 

\bibitem[Condon et al.(1991)]{Condon91} Condon, J. J., Huang, Z.-P., Yin, Q. F., \& Thuan, T. X. 1991, \apj, 378, 65

\bibitem[Condon(1992)]{Condon92} Condon, J. J. 1992, \araa, 30, 575

\bibitem[Cravens \& Dalgarno(1978)]{Cravens78} Cravens, T.~E., \& Dalgarno, A.\ 1978, \apj, 219, 750 

\bibitem[Crocker et al.(2011)]{Crocker11-Wild} Crocker, R.~M., Jones, D.~I., Aharonian, F., Law, C.~J., Melia, F., Oka, T., \& Ott, J.\ 2011, \mnras, 413, 763 

\bibitem[de Cea del Pozo et al.(2009)]{deCeaDelPozo09} de Cea del Pozo, E., Torres, D. F., \& Rodriguez Marrero, A. Y. 2009, \apj, 698, 1054 

\bibitem[Diehl et al.(2006)]{Diehl06} Diehl, R., Halloin, H., Kretschmer, K., et al.\ 2006, \aap, 449, 1025 

\bibitem[Domingo-Santamar\'ia \& Torres(2005)]{Domingo05} Domingo-Santamar\'ia, E. \& Torres, D. F. 2005, \aap, 444, 403

\bibitem[Groves et al.(2003)]{Groves03} Groves, B.~A., Cho, J., Dopita, M., \& Lazarian, A. 2003, Publ. Astron. Soc. Australia, 20, 252

\bibitem[Hales et al.(1991)]{Hales91} Hales, S.~E.~G., Mayer, C.~J., Warner, P.~J., \& Baldwin, J.~E.\ 1991, \mnras, 251, 46 

\bibitem[Heckman et al.(1990)]{Heckman90} Heckman, T.~M., Armus, L., \& Miley, G.~K.\ 1990, ApJS, 74, 833 

laxies
\bibitem[Israel \& Mahoney(1990)]{Israel90} Israel, F.~P., \& Mahoney, M.~J.\ 1990, \apj, 352, 30 

\bibitem[Kennicutt(1998)]{Kennicutt98} Kennicutt, R. C. 1998, \apj, 498, 541

\bibitem[Klein et al.(1988)]{Klein88} Klein, U., Wielebinski, R., \& Morsi, H.~W.\ 1988, \aap, 190, 41 

\bibitem[Kn{\"o}dlseder(1999)]{Knoedlseder99} Kn{\"o}dlseder, J.\ 1999, \apj, 510, 915 

\bibitem[Krumholz \& McKee(2005)]{Krumholz05} Krumholz, M.~R., \& McKee, C.~F.\ 2005, \apj, 630, 250 

\bibitem[Kulsrud \& Pearce(1969)]{Kulsrud69} Kulsrud, R., Pearce, W. P. 1969, \apj, 156, 445

\bibitem[Lacki et al.(2010)]{Lacki10-FRC1} Lacki, B.~C., Thompson, T.~A., \& Quataert, E.\ 2010, \apj, 717, 1 

\bibitem[Lacki et al.(2011)]{Lacki11-Obs} Lacki, B.~C., Thompson, T.~A., Quataert, E., Loeb, A., \& Waxman, E.\ 2011, \apj, 734, 107

\bibitem[Lacki(2012a)]{Lacki12-GRDRs} Lacki, B.~C.\ 2012a, arXiv:1204.2580 

\bibitem[Lacki(2012b)]{Lacki12-Al26} Lacki, B.~C.\ 2012b, arXiv:1204.2584 

\bibitem[Lacki(2012c)]{Lacki12-LowNu} Lacki, B.~C.\ 2012c, arXiv:1206.7100

\bibitem[Lacki(2013)]{Lacki12-FRCDerived} Lacki, B.~C. 2013, in prep.

\bibitem[Laing \& Peacock(1980)]{Laing80} Laing, R.~A., \& Peacock, J.~A.\ 1980, \mnras, 190, 903 

\bibitem[McCrady \& Graham(2007)]{McCrady07} McCrady, N., \& Graham, J.~R.\ 2007, \apj, 663, 844 

\bibitem[Mestel \& Spitzer(1956)]{Mestel56} Mestel, L., \& Spitzer, L., Jr.\ 1956, \mnras, 116, 503 

\bibitem[Meyer \& Clayton(2000)]{Meyer00} Meyer, B.~S., \& Clayton, D.~D.\ 2000, \ssr, 92, 133 

\bibitem[Murphy et al.(2008)]{Murphy08} Murphy, E.~J., Helou, G., Kenney, J.~D.~P., Armus, L., \& Braun, R.\ 2008, \apj, 678, 828 

\bibitem[Murphy et al.(2012)]{Murphy12} Murphy, E.~J., Porter, T.~A., Moskalenko, I.~V., Helou, G., \& Strong, A.~W.\ 2012, \apj, 750, 126 

\bibitem[Nord et al.(2006)]{Nord06} Nord, M.~E., Henning, P.~A., Rand, R.~J., Lazio, T.~J.~W., \& Kassim, N.~E.\ 2006, \aj, 132, 242 

\bibitem[Paglione et al.(1996)]{Paglione96} Paglione, T.~A.~D., Marscher, A.~P., Jackson, J.~M., \& Bertsch, D.~L.\ 1996, \apj, 460, 295 

\bibitem[Papadopoulos(2010)]{Papadopoulos10-CRDRs} Papadopoulos, P.~P.\ 2010, \apj, 720, 226 

\bibitem[Persic et al.(2008)]{Persic08} Persic, M., Rephaeli, Y., \& Arieli, Y. 2008, \aap, 486, 143

\bibitem[Persic \& Rephaeli(2012)]{Persic12} Persic, M., \& Rephaeli, Y.\ 2012, arXiv:1201.0369 

\bibitem[Rengarajan(2005)]{Rengarajan05} Rengarajan, T.~N.\ 2005, Proc. 29th Int. Cosmic Ray Conf. (Pune), 3

\bibitem[Rephaeli et al.(2010)]{Rephaeli10} Rephaeli, Y., Arieli, Y., \& Persic, M. 2010, \mnras, 401, 473

\bibitem[Robishaw et al.(2008)]{Robishaw08} Robishaw, T., Quataert, E., \& Heiles, C.\ 2008, \apj, 680, 981 

\bibitem[Rodriguez-Rico et al.(2004)]{RodriguezRico04} Rodriguez-Rico, C.~A., Viallefond, F., Zhao, J.-H., Goss, W.~M., \& Anantharamaiah, K.~R.\ 2004, \apj, 616, 783 

\bibitem[Roger et al.(1986)]{Roger86} Roger, R.~S., Costain, C.~H., \& Stewart, D.~I.\ 1986, \aaps, 65, 485 

\bibitem[Rybicki \& Lightman(1979)]{Rybicki79} Rybicki, G. B. \& Lightman, A. P. 1979, \emph{Radiative Processes in Astrophysics}, (New York: Wiley-VCH).

\bibitem[Stone et al.(1998)]{Stone98} Stone, J.~M., Ostriker, E.~C., \& Gammie, C.~F.\ 1998, \apjl, 508, L99 

\bibitem[Strickland \& Heckman(2007)]{Strickland07} Strickland, D.~K., \& Heckman, T.~M.\ 2007, \apj, 658, 258 

\bibitem[Strickland \& Heckman(2009)]{Strickland09} Strickland, D.~K., \& Heckman, T.~M.\ 2009, \apj, 697, 2030 

\bibitem[Strong \& Moskalenko(1998)]{Strong98} Strong, A.~W., \& Moskalenko, I.~V.\ 1998, \apj, 509, 212 

\bibitem[Suchkov et al.(1993)]{Suchkov93} Suchkov, A., Allen, R.~J., \& Heckman, T.~M.\ 1993, \apj, 413, 542 

\bibitem[Thompson et al.(2006)]{Thompson06} Thompson, T.~A., Quataert, E., Waxman, E., Murray, N., 
\& Martin, C.~L.\ 2006, \apj, 645, 186 

\bibitem[Thompson et al.(2007)]{Thompson07} Thompson, T. A., Quataert, E., Waxman, E. 2007, \apj, 654, 219

\bibitem[Torres(2004)]{Torres04} Torres, D. F. 2004, \apj, 617, 966

\bibitem[Torres et al.(2012)]{Torres12} Torres, D.~F., Cillis, A., Lacki, B., \& Rephaeli, Y.\ 2012, \mnras, 423, 822 

\bibitem[Umebayashi \& Nakano(2009)]{Umebayashi09} Umebayashi, T., \& Nakano, T.\ 2009, \apj, 690, 69 

\bibitem[V\"olk(1989)]{Volk89} V\"olk, H. J. 1989, \aap, 218, 67

\bibitem[Weiss et al.(2001)]{Weiss01} Wei{\ss}, A., Neininger, N., H{\"u}ttemeister, S., \& Klein, U.\ 2001, \aap, 365, 571 

\bibitem[Williams \& Bower(2010)]{Williams10} Williams, P.~K.~G., \& Bower, G.~C.\ 2010, \apj, 710, 1462 

\bibitem[Wills et al.(1997)]{Wills97} Wills, K.~A., Pedlar, A., Muxlow, T.~W.~B., \& Wilkinson, P.~N.\ 1997, \mnras, 291, 517 

\end{thebibliography}
\end{document}